\documentclass[a4paper,fleqn]{cas-dc}

\def\tsc#1{\csdef{#1}{\textsc{\lowercase{#1}}\xspace}}
\tsc{WGM}
\tsc{QE}
\usepackage{amssymb}
\usepackage{amsthm}
\usepackage{amsmath}

\usepackage{makeidx}  
\usepackage{graphicx}
\usepackage{amsfonts, amssymb, amsmath, cite,bm}
\usepackage{algorithm}
\usepackage{algorithmic}
\usepackage{color}
\usepackage{float}
\usepackage{orcidlink}
\usepackage{multirow}
\usepackage{multicol}
\newtheorem{thm}{Theorem}

\newtheorem{lem}[thm]{Lemma}

\newtheorem{defn}{Definition}
  
\newcommand{\eq}{\begin{equation}}
\newcommand{\nq}{\end{equation}}
\newcommand\eqa{\begin{eqnarray}}
\newcommand\nqa{\end{eqnarray}}

\newcommand\bess{\begin{eqnarray*}}
\newcommand\eess{\end{eqnarray*}}
\newcommand{\nn}{\nonumber}

\newcommand{\ALG}{}
\newenvironment{breakablealgorithm}
{
	\begin{center}
		\refstepcounter{algorithm}
		\hrule height.8pt depth0pt \kern2pt
		\renewcommand{\caption}[2][\relax]{
			{\raggedright\textbf{\ALG Algorithm~\thealgorithm} ##2\par}%
			\ifx\relax##1\relax 
			\addcontentsline{loa}{algorithm}{\protect\numberline{\thealgorithm}##2}%
			\else 
			\addcontentsline{loa}{algorithm}{\protect\numberline{\thealgorithm}##1}%
			\fi
			\kern2pt\hrule\kern2pt
		}
	}{
		\kern2pt\hrule\relax
	\end{center}
}

\begin{document}
\let\WriteBookmarks\relax
\def\floatpagepagefraction{1}
\def\textpagefraction{.001}

\shorttitle{The Restricted Inverse Optimal Value Problem}    

\shortauthors{Q. Zhang et~al.}  

\title [mode = title]{The Restricted Inverse Optimal Value Problem on Shortest Path on Trees under Weighted Bottleneck Hamming distance }  

\tnotemark[1] 

\tnotetext[1]{The work of Q. Zhang was supported by National Natural Science Foundation of China (1230012046).} 

\author[1]{Qiao Zhang \orcidlink{0009-0004-4723-2017} }\cormark[1]\ead{qiaozhang@cczu.edu.cn}
\cortext[1]{Corresponding author}
\author[2]{Xiao Li \orcidlink{0009-0003-4222-6577}}\ead{xli2576@wisc.edu}
\author[3]{Xiucui Guan \orcidlink{0000-0002-2653-1868}}\ead{xcguan@163.com}

\address[1]{Aliyun School of Big Data, Changzhou University, Changzhou 213164, Jiangsu, China }
\address[2]{Department of Mathematics, University of Wisconsin, Madison, WI 53706, USA  }
\address[3]{School of Mathematics, Southeast University, Nanjing 210096, Jiangsu, China}
%

















\begin{abstract}
We consider the Restricted Inverse Optimal Value Problem  on Shortest Path  on Trees (RIOVSPT$_{BH}$) under weighted bottleneck Hamming distance.
The problem aims to minimize the total cost under weighted bottleneck Hamming distance such that the length of the shortest root-leaf path of the tree is lower-bounded by a given value by adjusting the lengths of some edges. Additionally, the specified lower bound must correspond to the length of a particular root-leaf path.
Through careful analysis of the problem's structural properties, we develop an algorithm with $O(n\log n)$ time complexity to solve (RIOVSPT$_{BH}$). Furthermore, by removing the path-length constraint, we derive the Minimum Cost Shortest Path Interdiction Problem on Trees, for which we present an $O(n\log n)$ time algorithm that operates under weighted bottleneck Hamming distance. Extensive computational experiments demonstrate the efficiency and effectiveness of both algorithms.
\end{abstract}





\maketitle

\section{Introduction}

The inverse shortest path problem \textbf{(ISP)} has been a foundational topic in inverse optimization since its introduction by Burton and Toint\cite{BT-MP-ISP-1992}. 
Consider a weighted graph $G=(V,E,w)$ with a set of $s$-$t$ paths $P_j(j=0,1,\ldots,r-1)$, where $V$ and $E$ denote the sets of nodes and edges, respectively. The \textbf{(ISP)} seeks to minimally adjust the weight vector such that a prescribed path $P_0$ becomes the shortest path under some measurement criterion. For a vector $f$ and a path $P_j$, we define $f(P_j):=\sum_{e\in P_{j}}f(e)$.
Then the mathematical model for the problem \textbf{(ISP)} \cite{BT-MP-ISP-1992} can be stated as follows.
\begin{eqnarray}
&\min_{\bar w}&\|\bar w-w\| \nonumber\\
\textbf{(ISP)} &s.t.&  \bar w(P_0)\leq  \bar w(P_j),j=1,2,\cdots, r-1,\nn\\
&&\bar w(e)\geq 0, e\in E.\nonumber
\end{eqnarray}



A more challenging variant, the inverse optimal value problem on shortest paths \textbf{(IOVSP)}, has emerged as an important research direction, though it has received comparatively less attention than \textbf{(ISP)}. In \textbf{(IOVSP)}, rather than providing a desired solution a priori, the objective is to achieve a specified optimal value. Given a weighted graph $G=(V,E,w)$, the \textbf{(IOVSP)} problem seeks to adjust the weight vector while minimizing the total adjustment cost, such that the shortest path length equals a prescribed value $D$. Next is the mathematical formulation of \textbf{(IOVSP)}~\cite{CH-N-ISPL}:
\begin{eqnarray}
&\min_{\tilde w}&\|\tilde w-w\| \nonumber\\
\textbf{(IOVSP)} &\text{s.t.}& \min\limits_{0\leq j\leq r-1} \tilde w(P_j)=  D, \nonumber\\
&&\tilde w(e)\geq 0, \quad e\in E.\nonumber
\end{eqnarray}

Now we consider a restricted inverse optimal value problem on shortest path on
trees (denoted by \textbf{(RIOVSPT)}). Let $T=(V, E, w)$ be an edge-weighted tree rooted at $v_1$, where $V:=\{v_1, v_2, \cdots, v_n\}$ and  $E:=\{e_2, e_3, \cdots, e_{n}\}$ are the sets of nodes and edges, respectively. Let  $Y:=\{t_0,t_1, \cdots$ $,t_{r-1}\}$ be the set of leaves. Let $w(e)$, $l(e)$ and $u(e)$ be the  original weight, the lower and upper bounds of the adjusted weight for an edge $e\in E$, respectively,  where $l(e)\leq w(e)\leq u(e)$.  
Let $c(e)>0$ be a cost to ajust the weight of edge $e\in E$. Denoted by $P_{v_i, v_j}$ the unique path from $v_i$ to $v_j$ on the tree $T$ and record $P_k:=P_{v_1, t_k}$ as the ``root-leaf path" of the leaf node $t_k(k=0,1,2,\cdots,r-1)$ in brief.
The problem \textbf{(RIOVSPT)} aims at adjusting some weights of edges to minimize the total cost under some measurement such that the length of the shortest path of the tree is lower-bounded by a given value $D$, which is just the restriction of the length of the given path  $P_0$.  The mathematical model can be stated below.
\begin{eqnarray}
&\min_{\tilde w}&\|\tilde w-w\| \nonumber\\
\textbf{ (RIOVSPT) }&s.t.&\tilde w(P_k)\geq D, t_k\in Y, \label{eq-RIOVSPT}\\
&&  \tilde w(P_0)=D,\nonumber\\
&& l(e)\leq \tilde w(e)\leq u(e), e\in E.\nonumber
\end{eqnarray}

The feasible region of \textbf{(RIOVSPT)} can be characterized as the intersection of the feasible regions of \textbf{(ISP)} and \textbf{(IOVSP)} when restricted to tree structures. This structural relationship provides important insights into the problem's properties and solution space.

Zhang et al.\cite{ZhangQ22} proposed an $O(n^2)$ time algorithm to solve the problem \textbf{(RIOVSPT$_1$)} under weighted $l_1$ norm. Meanwhile, they also solved the problem \textbf{(RIOVSPT$_1$)} under unit $l_1$ norm in linear time. 

In a seminal contribution to the field, Zhang et al.~\cite{ZGP-LBIOVMST-JGO,ZGZ-IOVMST-OL} introduced the inverse optimal value problem on minimum spanning trees \textbf{(RIOVMST)}. This problem incorporates dual constraints: ensuring a specified spanning tree $T^0$ maintains its minimum spanning tree status while simultaneously achieving a prescribed weight value $K$. Their research under the unit $l_{\infty}$ norm for measuring adjustment costs yielded two strongly polynomial-time algorithms, both achieving $O(mn)$ time complexity---one for the unbounded case~\cite{ZGZ-IOVMST-OL} and another for the lower-bounded case~\cite{ZGP-LBIOVMST-JGO}, where $|V|:=n, |E|:=m$.

Building upon this foundation, Jia et al.~\cite{Jia2023} made several significant contributions. For the $l_\infty$ norm variant, they developed an efficient binary search method achieving $O(m^2 \log n)$ time complexity, while also proposing a more efficient $O(mn)$ algorithm specifically for the unit $l_\infty$ norm case. Their subsequent work~\cite{Jia2024} addressed the more challenging weighted $l_1$ norm variant, presenting an algorithm with time complexity $O(m^2 n^2 \log n \log(n c_{\max}))$, where $c_{\max}$ denotes the maximum element in the cost vector $c$.
Complementing these advances, Wang et al.~\cite{WGZ-CIOVMST-JOCO} introduced novel algorithmic techniques for the bounded \textbf{RIOVMST} problem under bottleneck Hamming distance, achieving an $O(mn \log m)$ time complexity.

\begin{figure}
    \centering
\includegraphics[width=0.7\linewidth]{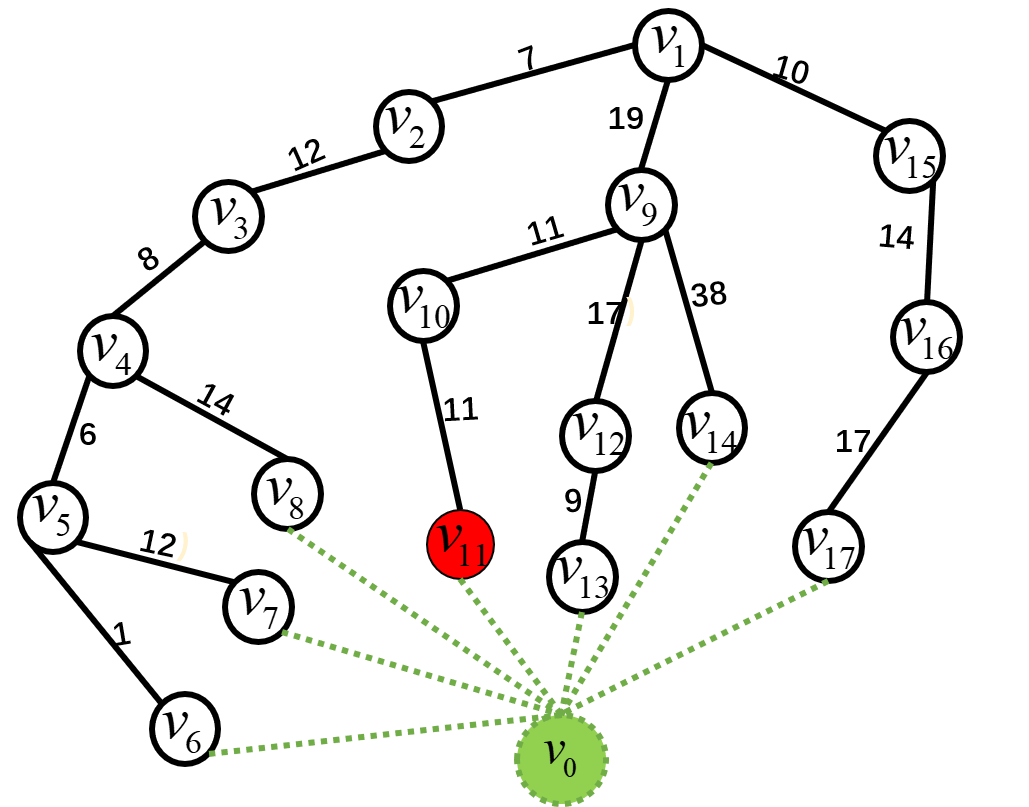}
    \caption{\cite{ZhangQ22}A special bandwidth network between the cities $v_1$ and
$v_0$ with prices on the links.}
    \label{Fig1}
\end{figure}

The practical significance of \textbf{(RIOVSPT)} is particularly evident in telecommunication bandwidth pricing, as highlighted by Paleologo and Takriti~\cite{PT-ANV-BT}. Consider a bandwidth provider pricing city-to-city links on a specialized bandwidth tree network (excluding node $v_0$) in \textbf{Figure.\ref{Fig1}}\cite{ZhangQ22}. Given an observed traded price $D$ for a bandwidth link between cities $v_1$ and $v_0$, the sum of prices along the cheapest path between these cities must equal $D$ to prevent arbitrage opportunities. Without this equilibrium, traders could exploit price differentials by purchasing the cheaper option (either the direct traded link or the sequence of links along the cheapest path) and selling the more expensive alternative.
In this context, suppose path $P_{11}$ through node $v_{11}$ represents the optimal route for emergency response scenarios, while alternative paths may be suboptimal due to construction budget constraints. To ensure reliable emergency communication between $v_1$ and $v_0$, the total price along path $P_{11}$ must equal $D$.

Zhang, Guan et al. \cite{ZhangQ20, ZhangQ21} studied the minimum cost shortest path interdiction problem by upgrading edges on trees (\textbf{MCSPIT}). In their work, they did not consider the restriction on the path \( P_0 \) as outlined in problem (\textbf{RIOVSPT}). The \textbf{(MCSPIT)} problem seeks to minimize the total cost of upgrading selected edges under a given measurement, ensuring that the length of the shortest path on the tree is bounded below by a specified value \( D \). The mathematical model for this problem can be formulated as follows.
\begin{eqnarray}
&\min_{\bar w}&  \|\bar w-w\|\nonumber\\
\textbf{(MCSPIT)}&s.t.&\bar w(P_k)\geq D, t_k\in Y, \label{eq-MCSPIT}\\
&& w(e)\leq \bar w(e)\leq u(e), e\in E.\nonumber
\end{eqnarray}

It is obvious that the optimal solution of the problem (\textbf{RIOVSPT}) is a feasible solution of the problem \textbf{(MCSPIT)}. 

Under the weighted $ l_1 $ norm, Zhang, Guan et al.\cite{ZhangQ20} proposed a primal-dual algorithms  in $ O(n^2) $ time for the problems \textbf{(MCSPIT$_1$)}. Under the weighted Hamming distance, they\cite{ZhangQ21} demonstrated that the  problem are  $\mathcal{NP}$-hard. While for the unit sum-Hamming distance,  they developed an  $ O(n^4\log n) $ time algorithm for the problem  \textbf{(MCSPIT$_{\bm{USH}}$)} by  employing a combination of dynamic programming .   Subsequently, with some local optimization, Yi et al. \cite{YiL22} improved upon this algorithm, achieving two reduced time complexity of  $ O(n^3 \log n) $ for this problem. 


when the weighted bottleneck Hamming distance is applied to $\|\cdot\|$ in models (\ref{eq-RIOVSPT})-(\ref{eq-MCSPIT}), the problems \textbf{(RIOVSPT$\bm{_{BH}}$)} and  \textbf{(MCSPIT$\bm{_{BH}}$)} under weighted bottleneck Hamming distance  can be formulated as follows. 
 \begin{eqnarray}
 &\min_{\tilde w}& C(\tilde w):=\max_{e\in E}c(e)H(\tilde w(e),w(e)) \nonumber\\
 \textbf{(RIOVSPT$\bm{_{BH}}$)} &s.t.&\tilde w(P_k)\geq D, t_k\in Y, \nonumber\\
 && \tilde w(P_0)=D,\label{eq-RIOVSPTBH}\\
 && l(e)\leq \tilde w(e)\leq u(e), e\in E.\nonumber
 \end{eqnarray}

\begin{eqnarray}
&\min_{\bar w}& M(\bar w):=\max_{e\in E}c(e)H(\bar w(e),w(e)) \nonumber\\
\textbf{(MCSPIT$\bm{_{BH}}$)}&s.t.&\bar w(P_k)\geq D, t_k\in Y, \label{eq-MCSPITBH}\\
&& w(e)\leq \bar w(e)\leq u(e), e\in E.\nonumber
\end{eqnarray}
where Hamming distance $H(\bar w(e),w(e))$ between $\bar w(e)$ and $w(e)$ is defined as $$H(\bar w(e),w(e))=\left\{\begin{array}{ll}1,&\bar w(e)\neq w(e),\\0,&\bar w(e)= w(e).\end{array}\right.$$

 
This paper is structured as follows: Section 2 presents our solution methodology for the Restricted Inverse Optimal Value Problem on Trees under weighted bottleneck Hamming distance (\textbf{RIOVSPT}$_{BH}$). Section 3 develops an algorithmic approach for the Minimum Cost Shortest Path Interdiction Problem (\textbf{MCSPIT}$_{BH}$). Section 4 evaluates the computational performance of both algorithms through extensive numerical experiments. Finally, Section 5 concludes the paper with a summary of our findings and directions for future research.

\section{Solve the problems (RIOVSPT$_{BH}$).}

For problem (RIOVSPT$_{BH}$) in the model (\ref{eq-RIOVSPTBH}), the optimal value must be one of the costs of edges. Notice that once an edge $e$ is adjusted, no matter how or in what ajusting amount, the relevant Hamming distance of $e$ is 1 and its adjusting cost is $c(e)$. Hence, suppose $\tilde c$ is an objective value. Then the edges with costs less than $\tilde c$ can be adjusted to maintain the objective value.

For convenience, we first sort the edges $e$ by their values of $c(e)$ 
   in non-decreasing order and removing any duplicates to produce the sequence $ \{\alpha_{n^*}\} $:    
\begin{equation} \label{eq:Sequence}
    c(e_{\alpha_1})  < c(e_{\alpha_2}) < \cdots < c(e_{\alpha_{n^*}}).
\end{equation}

		 Let $C_k:=c(e_{\alpha_k})$, $k=1,2,\cdots,{n^*}$. 

Then for any one cost $C_k$, the feasibility of the problem \textbf{(RIOVSPT$\bm{_{BH}}$)} can be judged in the following theorem, where $E_k := \{e \in E \mid c(e) \leq C_k\}$. Let $\mathcal{C} := \{C_k \mid k=1,2,\cdots,{n^*}\}$ be the set of costs of edges in the sequence $ \{\alpha_{n^*}\} $.

\begin{thm}\label{thm-fea=RIOVSPTBH}
The problem \textbf{(RIOVSPT$\bm{_{BH}}$)} has a feasible solution with objective value $C_k \in \mathcal{C}$ if and only if:
\begin{enumerate}
    \item $l(P_0 \cap E_k) + w(P_0 \setminus E_k) \leq D \leq u(P_i \cap E_k) + w(P_i \setminus E_k)$ for all $t_i \in Y$, and
    \item $l(P_0 \cap E_k) + w(P_0 \setminus E_k) \leq l(P_i \cap P_0 \cap E_k) \\
    + u(P_i \setminus P_0 \cap E_k) + w(P_i \setminus E_k)$ for all $t_i \in Y$.
\end{enumerate}
\end{thm}
\begin{proof}
 \textbf{Necessity.} If $\tilde{w}$ is a feasible solution of the problem \textbf{(RIOVSPT$\bm{_{BH}}$)}, then we have $l(e) \leq \tilde{w}(e) \leq u(e)$ for all $e \in E$. Moreover,
$$
u(P_i \cap E_k) + w(P_i \setminus E_k) \geq \tilde{w}(P_i) \geq D, \quad \forall t_i \in Y \setminus \{t_0\},
$$
and
$$\begin{aligned}
u(P_0 \cap E_k) + w(P_0 \setminus E_k) &\geq \tilde{w}(P_0) = D \\&\geq l(P_0 \cap E_k) + w(P_0 \setminus E_k).
\end{aligned}
$$
Hence, condition 1 holds.

Next, suppose that
$$\begin{aligned}
&l(P_0 \cap E_k) + w(P_0 \setminus E_k)\\& > l(P_{t^*} \cap P_0 \cap E_k) + u(P_{t^*} \setminus P_0 \cap E_k) + w(P_{t^*} \setminus E_k),\end{aligned}
$$
where
$$
t^* := \arg\min_{t_i \in Y} \Big(l(P_i \cap P_0 \cap E_k) + u(P_i \setminus P_0 \cap E_k) + w(P_i \setminus E_k)\Big).
$$

\textbf{Case (i):} If $P_{t^*} \cap P_0 = \emptyset$, then
$$\begin{aligned}
D &\geq l(P_0 \cap E_k) + w(P_0 \setminus E_k) \\&> u(P_{t^*} \cap E_k) + w(P_{t^*} \setminus E_k) \geq \bar{w}(P_{t^*}),\end{aligned}
$$
which contradicts with the feasibility of $\bar{w}$.

\textbf{Case (ii):} If $P_{t^*} \cap P_0 \neq \emptyset$, we have
\begin{eqnarray*}
   && u(P_{t^*} \setminus P_0 \cap E_k) + w(P_{t^*} \setminus P_0 \setminus E_k) < l(P_0 \cap E_k) \\&&+ w(P_0 \setminus E_k) - l(P_{t^*} \cap P_0 \cap E_k) - w(P_{t^*} \cap P_0 \setminus E_k)\\&=&l(P_0 \setminus P_{t^*} \cap E_k) + w(P_0 \setminus P_{t^*} \setminus E_k),
\end{eqnarray*}
and thus
$$\begin{aligned}
D &\leq \tilde{w}(P_{t^*}) = \tilde{w}(P_{t^*} \cap P_0) + \tilde{w}(P_{t^*} \setminus P_0) \\
&\leq \tilde{w}(P_{t^*} \cap P_0) + u(P_{t^*} \setminus P_0 \cap E_k) + w(P_{t^*} \setminus P_0 \setminus E_k) \\
&<\tilde{w}(P_{t^*} \cap P_0) + l(P_0 \setminus P_{t^*} \cap E_k) + w(P_0 \setminus P_{t^*} \setminus E_k) \\
&\leq \tilde{w}(P_0),\end{aligned}
$$
which contradicts with $\tilde{w}(P_0) = D$. Hence, condition 2 holds.

\textbf{Sufficiency.} Suppose the conditions 1-2 hold. Then we have 
\begin{eqnarray*}
&& l(P_0 \cap E_k) + w(P_0 \setminus E_k) \leq D\\&\leq&\min_{t_i \in Y}\big(u(P_i \cap E_k) + w(P_i \setminus E_k)\big) \\&\leq& u(P_0 \cap E_k) + w(P_0 \setminus E_k).   
\end{eqnarray*}

Let $ e_j = (v_i, v_j) \in P_0 \cap E_k $ be the first edge on $ P_0 \cap E_k $ from $ v_1 $ satisfying $$ 
\begin{aligned}
    & u(P_{v_1, v_i} \cap E_k) + w(P_{v_1, v_i} \setminus E_k) + l(P_{v_i, t_0} \cap E_k)\\ &+ w(P_{v_i, t_0} \setminus E_k) 
      \leq D\leq u(P_{v_1, v_j} \cap E_k) + w(P_{v_1, v_j} \setminus E_k)\\& + l(P_{v_j, t_0} \cap E_k) + w(P_{v_j, t_0} \setminus E_k).
\end{aligned}
$$
Notice that the special cases when $ v_i = v_1 $ with 
$$\begin{aligned}
&~~~~~u(P_{v_1, v_i} \cap E_k) + w(P_{v_1, v_i} \setminus E_k) \\&= u(P_{v_1, v_1} \cap E_k) + w(P_{v_1, v_1} \setminus E_k) = 0\end{aligned}
$$
and when $ v_j = t_0 $ with 
$$\begin{aligned}
&~~~~~l(P_{v_j, t_0} \cap E_k) + w(P_{v_j, t_0} \setminus E_k) \\&= l(P_{t_0, t_0} \cap E_k) + w(P_{t_0, t_0} \setminus E_k) = 0\end{aligned}
$$
have been considered. 

Then we can construct
\begin{equation} \label{fs-RIOVSPTBH}
   \tilde{w}(e) := 
\begin{cases} 
    u(e), & e \in E \setminus P_{v_i, t_0} \cap E_k, \\
    l(e), & e \in P_{v_j, t_0} \cap E_k, \\
   \delta, & e = e_j = (v_i, v_j), \\
    w(e), & \text{otherwise},
\end{cases}
\end{equation}
where 
$ 
\delta = D - u(P_{v_1, v_i} \cap E_k) - w(P_{v_1, v_i} \setminus E_k) $ 

$~~~~~~~~~~~$ $ - l(P_{v_j, t_0} \cap E_k) - w(P_{v_j, t_0} \setminus E_k).
$

Next, we show that $ \bar{w} $ is a feasible solution of the problem \textbf{(RIOVSPT$\bm{_{BH}}$)}.

\textbf{ (i)} For $ e \in E \setminus \{e_j\} $, we  have
$l(e) \leq \tilde{w}(e) \leq u(e). $ 


For $ e_j = (v_i, v_j) \in E_k $, we have 
$$
\begin{aligned}
    & u(P_{v_1, v_i} \cap E_k) + w(P_{v_1, v_i} \setminus E_k) + l(v_i, v_j) + l(P_{v_j, t_0} \cap E_k)\\&~~~~ + w(P_{v_j, t_0} \setminus E_k) = u(P_{v_1, v_i} \cap E_k) + w(P_{v_1, v_i} \setminus E_k) \\ &~~~~+ l(P_{v_i, t_0} \cap E_k) + w(P_{v_i, t_0} \setminus E_k)
    \leq D \leq u(P_{v_1, v_j} \cap E_k)  \\&~~~~+ w(P_{v_1, v_j} \setminus E_k) +l(P_{v_j, t_0} \cap E_k) + w(P_{v_j, t_0} \setminus E_k) 
   \\& = u(P_{v_1, v_i} \cap E_k)+ w(P_{v_1, v_i} \setminus E_k)  + u(v_i, v_j) \\&~~~~+ l(P_{v_j, t_0} \cap E_k) + w(P_{v_j, t_0} \setminus E_k).
\end{aligned}
$$
Thus, we have 
$ 
l(e_j) \leq \delta = \tilde{w}(e_j) \leq u(e_j).
$

\textbf{ (ii)} Let $Y^1:=\{t'\in Y|E(P_{t'}\cap P_0)=\emptyset\}$ be the set of the leaf nodes whose root-leaf paths have no intersecting edges with the path $P_0$. 
    
    For ${t'}\in Y^1,$ $\bar w(P_{t'})= u(P_{t'}\cap E_k)+w(P_{t'}\backslash E_k)\geq D$.
	$$\begin{aligned}
		&\tilde{w}(P_0)= \tilde w(P_{v_1,v_i})+  \tilde w(P_{v_j,t_0})+ \tilde w(v_i,v_j)\\ &=u(P_{v_1,v_i}\cap E_k)+w(P_{v_1,v_i}\backslash E_k) +l(P_{v_j,t_0}\cap E_k)\\&~~~~+w(P_{v_j,t_0}\backslash E_k)+\Big(D-u(P_{v_1,v_i}\cap E_k)-w(P_{v_1,v_i}\backslash E_k)\\&~~~~-l(P_{v_j,t_0}\cap E_k)-w(P_{v_j,t_0}\backslash E_k)\Big)=D.
	\end{aligned}$$
    
	For ${t'}\in Y\backslash (Y^1\cup \{t_0\})$, let $v^*_t\in P_{t'}\cap P_0$ be the last intersection node  on $P_{t'}\cap P_0$. 
    
	\textbf{(a)} If $v^*_t\in P_{v_1,v_i}$, then
	$\bar w(P_{t'})=u(P_{t'}\cap E_k)+w(P_{t'}\backslash E_k)\geq D.$
    
	\textbf{(b)} If $v^*_t=v_j$, we have$$ \begin{aligned}
	    &l(P_{v_1,v_j}\cap E_k )+w(P_{v_1,v_j}\backslash E_k)+l(P_{v_j,t_0}\cap E_k)\\&~~~~+w(P_{v_j,t_0}\backslash E_k)=l(P_0\cap E_k)+w(P_0\backslash E_k)\\&\leq \min_{t'\in Y}(l(P_{t'}\cap P_0\cap E_k)+u(P_{t'}\backslash P_0\cap E_k)+w(P_{t'}\backslash E_k))\\&\leq l(P_{v_1,v^*_t}\cap E_k)+u(P_{v^*_t,{t'}}\cap E_k)+w(P_{t'}\backslash E_k)\\&=l(P_{v_1,v_j}\cap E_k)+w(P_{v_1,v_j}\backslash E_k)+u(P_{v_j,{t'}}\cap E_k)\\&~~~~+w(P_{v_j,{t'}}\backslash E_k)
	\end{aligned}$$ 
    and thus 
      $$l(P_{v_j,t_0}\cap E_k)+w(P_{v_j,t_0}\backslash E_k)\leq u(P_{v_j,{t'}}\cap E_k)+w(P_{v_j,{t'}}\backslash E_k).$$ Then
	$$\begin{aligned}
		&\tilde w(P_{t'})=  \tilde w(P_{v_1,v_i})+ \tilde w(P_{v_j,{t'}})+\tilde w(v_i,v_j)\\&=u(P_{v_1,v_i}\cap E_k)+w(P_{v_1,v_i}\backslash E_k)+u(P_{v_j,{t'}}\cap E_k)\\&~~~~+w(P_{v_j,{t'}}\backslash E_k)+\Big(D-u(P_{v_1,v_i}\cap E_k)-w(P_{v_1,v_i}\backslash E_k)\\&~~~~-l(P_{v_j,t_0}\cap E_k)-w(P_{v_j,t_0}\backslash E_k)\Big)\geq D.
	\end{aligned}$$
	\textbf{(c)} If $v^*_t\in P_{v_j,t_0}\backslash \{v_j\}$, we have $$\begin{aligned}
	    &l(P_{v_1,v_j}\cap E_k)+w(P_{v_1,v_j}\backslash E_k)+l(P_{v_j,t_0}\cap E_k)\\&~~~~+w(P_{v_j,t_0}\backslash E_k)=l(P_0\cap E_k)+w(P_0\backslash E_k)\\&\leq \min_{t'\in Y}(l(P_{t'}\cap P_0\cap E_k)+u(P_{t'}\backslash P_0\cap E_k)+w(P_{t'}\backslash E_k))\\&\leq l(P_{v_1,v_j}\cap E_k)+w(P_{v_1,v_j}\backslash E_k)+l(P_{v_j,v^*_t}\cap E_k)\\&~~~~+u(P_{v^*_t,{t'}}\cap E_k)+w(P_{v_j,{t'}}\backslash E_k)
	\end{aligned} $$
    and thus $l(P_{v_j,t_0}\cap E_k)+w(P_{v_j,t_0}\backslash E_k)\leq l(P_{v_j,v^*_t})\\
    +u(P_{v^*_t,{t'}}\cap E_k)+w(P_{v_j,{t'}}\backslash E_k).$  Then we have
	$$\begin{aligned}
		&\tilde w(P_{t'})=  \tilde w(P_{v_1,v_i})+\tilde w(P_{v_j,v^*_t})+ \tilde w(P_{v^*_t,{t'}})+\tilde w(v_i,v_j)\\
		&=u(P_{v_1,v_i}\cap E_k)+w(P_{v_1,v_i}\backslash E_k)+l(P_{v_j,v^*_t}\cap E_k)\\&~~~~+u(P_{v^*_t,{t'}}\cap E_k)+w(P_{v_j,{t'}}\backslash E_k)+\Big(D-u(P_{v_1,v_i}\cap E_k)\\&~~~~-w(P_{v_1,v_i}\backslash E_k)-l(P_{v_j,t_0}\cap E_k)-w(P_{v_j,t_0}\backslash E_k)\Big)\geq D.
	\end{aligned}$$
	Therefore, the problem \textbf{(RIOVSPT$\bm{_{BH}}$)} is feasible. 
\end{proof}

For convenience, we intoduce the following definition.
\begin{defn}
    If a cost $\tilde c\in \mathcal{C}$ satisfies the conditions 1-2 in \textbf{Theorem \ref{thm-fea=RIOVSPTBH}}, we call $\tilde c$  feasible. Otherwise, we call it infeasible.
\end{defn}
  
Then we can solve  the problem \textbf{ (RIOVSPT$\bm{_{BH}}$)} by finding the minimum feasible cost in the sequence $ \{\alpha_{n^*}\} $. We can obviously have the following lemma.
\begin{lem}
      Suppose $\tilde c,\hat c\in \mathcal{C}$ and  $\hat c\geq \tilde c $.  If $\tilde c$ is feasible, so is $\hat c$. If $\hat c$ is infeasible, so is $\tilde c$. 
\end{lem}

 If the maximum cost $C_{n^*}$ in $\mathcal{C}$ is infeasible, we have the following theorem. 
\begin{thm}
    If  $C_{n^*}$ is infeasible, then the problem \textbf{ (RIOVSP-T$\bm{_{BH}}$)} is infeasible. 
\end{thm}

Hence, for the feasible problem, we can search for the minimum feasible cost by a binary search method based on sorting the costs on $\mathcal{C}$ ascendingly.Then a relevant optimal solution can be constructed as formula (\ref{fs-RIOVSPTBH}). 

Based on the analysis above, we propose \textbf{Algorithm \ref{alg-RIOVSPTBH}} to solve  the problem \textbf{ (RIOVSPT$\bm{_{BH}}$)} and followed by the corresponding time complexity analysis. 

\begin{breakablealgorithm}
	\renewcommand{\thealgorithm}{1}
	\caption{$[\tilde w, C(\tilde w)]$:=\textbf{RIOVSPT}$_{BH}(T, w,l, u, c, D,t_0)$  }
	\label{alg-RIOVSPTBH}
	\begin{algorithmic}[1]
		
		\REQUIRE  The tree $T(V,E)$ rooted at a root node $v_1$ with the set $Y$ of leaf nodes; the cost vector $c$, three weight vectors $w,l,u$ and the given leaf node $t_0$ and a given value $D$.
		\ENSURE  An optimal solution $\tilde w$ and its relevant objective value $C(\tilde w)$.
        
\IF {$\max_{e\in E}c(e)$ is infeasible}
\RETURN `` The problem is infeasible!"
\ELSE
		\STATE Sort the values of $c(e)$ 
   in non-decreasing order and removing any duplicates to produce the sequence $ \{\alpha_{n^*}\} $:    
\begin{equation*} 
    c(e_{\alpha_1})  < c(e_{\alpha_2}) < \cdots < c(e_{\alpha_{n^*}}).
\end{equation*}

		\STATE Let $C_j:=c(e_{\alpha_j})$, $j=1,2,\cdots,{n^*}$. 
        
            \STATE Initialization: $a := 1, b := n^*$, $k^* := 0, k := \lfloor\frac{a+b}2\rfloor$.
          
            \WHILE{$k^* =0$} 
             \IF {$a=b$ or $b=a+1$}
             \STATE $k^* := b$, \textbf{break.}
             
           \ENDIF
             \IF {$C_k$ is feasible}
              
             \STATE Update $b := k, k := \lfloor\frac{a+b}2\rfloor$.
            \ELSE \STATE  Update $a := k, k := \lfloor\frac{a+b}2\rfloor$.

            \ENDIF
            \ENDWHILE
		\RETURN $\tilde w$ as formula (\ref{fs-RIOVSPTBH}), $C(\tilde w):=C_{k^*}$.
\ENDIF		
	\end{algorithmic}
\end{breakablealgorithm}
\begin{thm}
The problem \textbf{ (RIOVSPT$\bm{_{BH}}$)} can be solved by \textbf{Algorithm 1} in $O(n\log n)$ time.
\end{thm}
\begin{proof}
   The main calculations are in \textbf{Lines 4-18} in \textbf{Algorithm \ref{alg-RIOVSPTBH}}. Sorting in \textbf{Line 4} costs $O(n\log n)$ time and there are $O(\log n )$ iterations in  the binary search method in \textbf{Line 7-16}, where it takes $O(n)$ time to check the feasibility of $C_k$ in \textbf{Line 11} in each iteration.  Hence, the theorem holds.
\end{proof}

\section{Solve the problem \textbf{(MCSPIT$\bm{_{BH}}$)}}

We next solve the problem \textbf{(MCSPIT$\bm{_{BH}}$)} based on solving the problem \textbf{(MSPIT$\bm{_{BH}}$)}, which aims to maximize the root-leaf shortest path of a tree while ensuring that the total cost of upgrading selected edges under weighted bottle-neck Hamming distance  does not exceed a given value $M$. The mathematical model for this problem can be formulated as follows.
 \begin{eqnarray}
&\max_{\hat w}& D(\hat  w):=\min_{t_k\in Y}\hat w(P_k) \label{eq-MSPITBH}\\
\textbf{(MSPIT$\bm{_{BH}}$)}&s.t.&\max_{e\in E}c(e)H(\hat w(e),w(e)) \leq M, \nonumber\\
&& w(e)\leq \hat w(e)\leq u(e), e\in E.\nonumber
\end{eqnarray}
 
Here is the property of the optimal solutions of the two problems.

\begin{thm}[\cite{ZhangQ21},Theorem 4]
In model (\ref{eq-MCSPITBH})-(\ref{eq-MSPITBH}), for $e\in E$, if the edge length vector $\tilde w$ is optimal, so is the edge length vector $w^*$ defined as follows.
$$w^*(e)=\left\{
\begin{array}{ll}
w(e), &\text{if  }\tilde {w}(e)=w(e)\\
u(e), &\text{if  }\tilde {w}(e)\neq{w(e)}.
\end{array}\right.$$
\end{thm}

Hence, in the problems \textbf{(MSPIT$\bm{_{BH}}$)} and \textbf{(MCSPIT$\bm{_{BH}}$)}, if one edge is upgraded,then we can upgrade it to its upper bound.

\begin{thm}
	The solution \begin{eqnarray}\hat w(e)=\left\{\begin{array}{ll}u(e),&c(e)\leq M\\w(e),&c(e)>M\end{array}\right.\label{os-MSPITBH}\end{eqnarray} is an optimal solution of the problem \textbf{(MSPIT$\bm{_{BH}}$)}.
\end{thm}
\begin{proof}
\textbf{(1)} For any edge $ e \in E $ with $ c(e) \leq M $, we have $ w(e) \leq \hat{w}(e) = u(e) $, and $ c(e)H(\hat{w}(e), w(e)) = c(e) \leq M $. For any edge $ e \in E $ with $ c(e) > M $, we have $ \hat{w}(e) = w(e) \leq u(e) $ and $ c(e)H(\hat{w}(e), w(e)) = 0 \leq M $. Therefore,  
$$
\max_{e \in E} c(e)H(\hat{w}(e), w(e)) \leq M.
$$
This implies that $ \hat{w} $ is a feasible solution of the problem \textbf{(MSPIT$\bm{_{BH}}$)}.

\textbf{(2)} Suppose $ \hat{w} $ is not optimal, and let $ w^* $ be an optimal solution. Then,  
$
D(w^*) = w^*(P_{t^*}) > D(\hat{w}) = \hat{w}(P_{\hat {t}}).
$

\textbf{Case (i):} If $ t^* = \hat{t} $, then there exists an edge $ \hat{e} \in P_{t^*} $ such that $ u(\hat{e}) \geq w^*(\hat{e}) > \hat{w}(\hat{e}) = w(\hat{e}) $ as $ w^*(P_{t^*}) > \hat{w}(P_{t^*}) $. In this case,  
$
c(\hat{e})H(w^*(\hat{e}), w(\hat{e})) = c(\hat{e}) > M,
$
which contradicts with the feasibility of $ w^* $.

\textbf{Case (ii):} If $ t^* \neq \hat{t} $, then $ w^*(P_{\hat{t}}) \geq w^*(P_{t^*}) > \hat{w}(P_{\hat{t}}) \leq \hat{w}(P_{t^*}) $. Consequently, there exists an edge $ \hat{e} \in P_{\hat{t}} $ such that $ u(\hat{e}) \geq w^*(\hat{e}) > \hat{w}(\hat{e}) = w(\hat{e}) $. Similarly,  
$$
c(\hat{e})H(w^*(\hat{e}), w(\hat{e})) = c(\hat{e}) > M,
$$
which again contradicts with the feasibility of $ w^* $.

Thus, in both cases, $ \hat{w} $ must be an optimal solution to the problem \textbf{(MSPIT$\bm{_{BH}}$)}.
\end{proof}

\begin{thm}
The problem  \textbf{(MSPIT$\bm{_{BH}}$)} can be solved by formla (\ref{os-MSPITBH})  in $ O(n) $ time.
\end{thm}

We next address the problem \textbf{(MCSPIT$\bm{_{BH}}$)}. From model (\ref{eq-MCSPITBH}), the objective value is one of the costs of edges. For any one cost $\tilde c\in \mathcal{C} := \{C_k \mid k=1,2,\cdots,{n^*}\}$,  we consider the length  of shortest path based on this cost $\tilde c$, which is just the problem \textbf{(MSPIT$\bm{_{BH}}$)} with $M=\tilde c.$
\begin{lem}
    If $\hat c,\tilde c\in \mathcal{C}$ and $\hat c>\tilde c$,  we can obtain $ \hat{w}$ with $\hat c$ and $\tilde{w}$ with $\tilde c$ from formula (\ref{os-MSPITBH}).  
    Then $D(\hat{w})\geq D(\tilde{w})$, where $D(\hat{w}):=\min_{t_k\in Y}\hat w(P_k)$. 
\end{lem}

By a binary search method based on the sequence $\{\alpha_{n^*}\}$ in (\ref{eq:Sequence}), we can determine the minimum cost on $\mathcal{C}$ whose corresponding  length of  shortest path is  no less than $D$.  Then the optimal objective value of  the problem \textbf{(MCSPIT$\bm{_{BH}}$)} can be obtained, so does its optimal solution.

The process begins by the ascending sequence $\{\alpha_{n^*}\} $ in (\ref{eq:Sequence}). 
For any $ C_j= c(e_{\alpha_j}) $, we can obtain an optimal solution $ \hat{w}^j $ of the problem \textbf{(MSPIT$\bm{_{BH}}$)} by replacing $M$ with $C_j$ in (\ref{os-MSPITBH}) as well as its corresponding shortest root-leaf distance $D_j:=D(\hat{w}^j ):=\min_{t_i\in Y}\hat{w}^j(P_i)$.  
Since $\{ C_j \}$ is increasing,  the output $ D_j $ obviously is  non-decreasing.

Specifically, for $ C_{n^*}  $ and $ C_1 $, from formula (\ref{os-MSPITBH}), we can achieve the relevant optimal solutions $\hat{w}^{n^*}$ with  its shortest root-leaf distance $ D_{n^*}:=D(\hat{w}^{n^*})$ and  $\hat{w}^1$ with $D_1:=D(\hat{w}^1)$,respectively. 

 Based on above analysis, we derive the following straightforward lemmas:

\begin{lem}
The optimal objective value of the problem \textbf{(MCSPIT$\bm{_{BH}}$)} lies within the range $ [0, C_{n^*}] $.
\end{lem}

\begin{lem}\label{lem-MCSPITinfty-feasible}
If $ \bar {w} $ is an optimal solution to the problem \textbf{(MCSPIT$\bm{_{BH}}$)}, then the relevant shortest root-leaf distance $ D(\bar{w}) $ satisfies $ D(w) \leq D(\bar{w}) \leq D_{n^*} $.
\end{lem}

By \textbf{Lemma \ref{lem-MCSPITinfty-feasible}}, the problem \textbf{(MCSPIT$\bm{_{BH}}$)} is infeasible for $D > D_{n^*}$, as all the weights of edges have been upgraded to their maximum capacity. Conversely, for $D \le D(w)$, the optimal solution is $w$ with an optimal value $0$, since the weight vector $w$ already satisfies the feasibility conditions. Consequently, we can state the following theorems.

\begin{thm}
	If $D> D_{n^*}$, the problem \textbf{(MCSPIT$\bm{_{BH}}$)} is infeasible.
\end{thm}

\begin{thm}
     If $D \le D(w)$, then $w$ is an optimal solution of the problem \textbf{(MCSPIT$\bm{_{BH}}$)}   and the optimal value is $0$.
\end{thm}

Then we consider the problem \textbf{(MCSPIT$\bm{_{BH}}$)} when $D(w)<D\leq D_{n^*}$. 

To be specific, based on the ascending sequence in \eqref{eq:Sequence} and a binary search method, we can determine the index $k$ satisfying $D_k < D \leq D_{k+1}$ and then the corresponding $C_{k+1}$ is just the minimum cost to satisfy the constraints of the problem \textbf{(MCSPIT$\bm{_{BH}}$)}. 



Based on the analysis above, we have the following   \textbf{Algorithm \ref{alg-MCSPITBH}} to solve the problem \textbf{(MCSPIT$\bm{_{BH}}$)}.

\begin{breakablealgorithm}
	\renewcommand{\thealgorithm}{2}
	\caption{$[\bar w, M^*]:=\textbf{MCSPIT}_{\bm{BH}}(T, w, u, c, D)$}
	\label{alg-MCSPITBH}
	\begin{algorithmic}[1]
		\REQUIRE  The tree $T(V,E)$ rooted at a root node $v_1$ with the set $Y$ of leaf nodes; the cost vector $c$, two weight vectors $w,u$ and a given value $D$.
		\ENSURE  An optimal solution $\bar w$ and its relevant objective value $M^*$.
		\STATE Sort the edges $e$ by their values of $c(e)$ in ascending order(remove the repeating elements and suppose there are  $n^*$ different values). Renumbered as follows: \\
		$c(e_{\alpha_1})<c(e_{\alpha_2})< \cdots< c(e_{\alpha_{n^*}}).$ 
		\STATE Let $C_j:=c(e_{\alpha_j})$, $j=1,2,\cdots,{n^*}$.
        \STATE$D(w):=\min_{t_k\in Y}w(P_k)$.\\
        Obtain $\hat w^{n^*}$ with $C_{n^*}$ from formula (\ref{os-MSPITBH}) and its relevant length of shortest root-leaf distance $D_{n^*}:=D(\hat w^{n^*})$.
		
         Obtain $\hat w^1$ with $C_1$ from formula (\ref{os-MSPITBH}) and its relevant length of shortest root-leaf distance $D_1:=D(\hat w^1)$.
        \IF{$D\leq D(w)$}
        \RETURN ($ w$, $0$).
		\ELSIF {$D> D_{n^*}$}
		\RETURN ``the problem is infeasible!"
		\ELSE
            \STATE Initialization: $a := 1, b := n^*$, $k^* := 0, k := \lfloor\frac{a+b}2\rfloor$.
            
            \WHILE{$k^* = 0$}
            \IF{$a = b$}
            \STATE $k^* := a$,  \textbf{break.}
            \ENDIF
            \STATE Obtain $\hat w^k$ with $C_k$ and its relevant length of shortest root-leaf distance $D_k :=D(\hat w^k)$.\\
          Obtain $\hat w^{k+1}$ with $ C_{k+1}$ and its relevant length of shortest root-leaf distance $D_{k+1} := D(\hat w^{k+1})$.
            \IF{$ D_{k+1} < D$}
            \STATE Update $a := k, k := \lfloor\frac{a+b}2\rfloor$.
            \ELSIF{$D_k > D$}
            \STATE Update $b := k, k := \lfloor\frac{a+b}2\rfloor$.
            \ELSIF{$D_{k}<D \le D_{k+1}$}
            \STATE $k^* := k+1$.
            \ENDIF
            \ENDWHILE
		\RETURN $\bar w:=w^{k^*}, M^*:=C_{k^*}$.
		\ENDIF
	\end{algorithmic}
\end{breakablealgorithm}
 The main calculations in \textbf{Algorithm \ref{alg-MCSPITBH}} are the cycles of the binary search method, where two \textbf{(MSPIT$\bm{_{BH}}$)} problems are solved in each iteration. Hence, 
\begin{thm}
The problem \textbf{(MCSPIT$\bm{_{BH}}$)} can be solved by \textbf{Algorithm \ref{alg-MCSPITBH}} in $O(n\log n)$ time.
\end{thm}

\section{Numerical experiments}

\subsection{An example to show the process of algorithms}
For the better understanding of \textbf{Algorithm \ref{alg-RIOVSPTBH}} and \textbf{Algorithm \ref{alg-MCSPITBH}}, Example 1 is given to show the detailed computing process.

\noindent\textbf{Example 1.} As is illustrated in  \textbf{Figure.\ref{Fig1}}, a special bandwidth network  between the cities $v_1$ and $v_0$ is given with prices on the links, which is a tree network(excluding the node $v_0$) with an observed traded price $D:=39$. Let $V:=\{v_1,v_2,\cdots,v_{17}\}$, $E:=\{e_2,e_3,\cdots,e_{17}\}$. Let the initial,lower and upper bounds vectors of the costs  are  $$\begin{aligned}
   &w:=(7, 12,     8,     6,    1,    12,      14,    19,    11,    11,    17,    9,    38,    10,   14,     17),\\&
   l: =(3 , 7 ,     4,    3,    0 ,    6 ,    4 ,   16 ,    10  ,   3,    13 ,    5 ,   37,     9,     8 ,    9),\\&
  u: =(9 , 18,    12,   9,     7,    14 ,   16 ,   22,    20 ,   14,    26,    12 ,   48,   14,    16 ,   20),
\end{aligned}$$  \text{respectively}.The penalty vector of adjusting the costs between cities is $$c:=(7,2,4,14,3,13,9,8, 2,7,2,1,15,12,13,14).$$

\textbf{(1)} Suppose path $P_{11}$ through node $v_{11}$ is the optimal route for emergency response scenarios. Try to determine a pricing scheme to prevent arbitrage opportunities and also to ensure reliable emergency communication.

\textbf{(2)} Try to determine a pricing scheme just to prevent arbitrage opportunities. 

We first show the detailed process to solve the  problem \textbf{(1)} by \textbf{Algorithm \ref{alg-RIOVSPTBH}}.

\textbf{In Lines 1-2:} $\max_{e\in E}c(e)=15,$ $E'=\{e\in E|c(e)\leq 15\}=E$. Then in \textbf{Theorem \ref{thm-fea=RIOVSPTBH}}, on the one hand, $l(P_0 \cap E') + w(P_0 \setminus E') =l(P_0) =29\leq D=39$ and $ \min_{t_k\in Y}u(P_k)=u(P_{v_{17}})=50\geq D:=40$, then  for all $t_k \in Y,  u(P_k \cap E') + w(P_k \setminus E')=P_k)\geq D:=39.$ On the other hand, 
$l(P_0 \cap E') + w(P_0 \setminus E') =l(P_0 )=29\leq \min_{t_k\in Y }\Big(l(P_k \cap P_0\cap E')+ u(P_k \setminus P_0 \cap E') + w(P_k \setminus E')\Big)=\min_{t_k\in Y }\Big(l(P_k \cap P_0)+ u(P_k \setminus P_0 ) \Big)=l(P_0)=29\leq l(P_k \cap P_0 \cap E')+ u(P_k \setminus P_0 \cap E') + w(P_k \setminus E')$ for all $t_k \in Y$. Hence, $\max_{e\in E}c(e)=15$ is feasible. 

\textbf{In Lines 4-5:} Construct the sequence $\{\alpha_{n^*}\}: C_1=c(e_{13})=1<C_2=c(e_{10})=2<C_3=c(e_{6})=3<C_4=c(e_{4})=4<C_5=c(e_{2})=7<C_6=c(e_{9})=8<C_7=c(e_{8})=9<C_8=c(e_{15})=12<C_9=c(e_{7})=13<C_{10}=c(e_{5})=14<C_{11}=c(e_{14})=15. $

\textbf{In Lines 7-16:a binary search method.}  \\
Initialization: $a:=1,b:=n^*=11,k^*:=0,k:=6$.

\textbf{In the first iteration}, $C_6=8, E_6=\{e_2,e_3,e_4,e_6,e_9,$ $e_{10},e_{11},e_{12},e_{13}\}$. $l(P_0 \cap E_6) + w(P_0 \setminus E_6)= l(P_0 \cap E_6)=29\leq D:=39\leq  \min_{t_k\in Y}\big( u(P_k \cap E_6) + w(P_k \setminus E_6)\big)=u(P_{v_{17}} \cap E_6) + w(P_{v_{17}} \setminus E_6)=41,$ $l(P_0 \cap E_6) + w(P_0 \setminus E_6)=29 \leq \min_{t_k\in Y}\big(l(P_k \cap P_0 \cap E_6)+ u(P_k \setminus P_0 \cap E_6) + w(P_k \setminus E_6)\big)=l(P_0)=29$.Hence, $C_6=8$ is feasible. Then update $b:=6,k:=3.$

\textbf{In the second iteration}, $C_3=3,E_3=\{e_3,e_6,e_{10},e_{12}\}.$  $l(P_0 \cap E_3) + w(P_0 \setminus E_3)=40>D:=40$. Hence, $C_3=3$ is infeasible. Then update $a:=3,k:=4.$

\textbf{In the third iteration}, $C_4=4,E_4=\{e_3,e_4,e_6,e_{10},e_{12},$ $e_{13}\}.$  $l(P_0 \cap E_4) + w(P_0 \setminus E_4)=40> D:=39$. Hence, $C_4=4$ is infeasible. Then update $a:=4,k:=5.$  

\textbf{In the fourth iteration}, $C_5=7, E_5=\{e_2,e_3,e_4,e_6,e_{10},$ $e_{11},e_{12},e_{13}\}$. $l(P_0 \cap E_5) + w(P_0 \setminus E_5)= l(P_0 \cap E_6)=32\leq D:=39\leq  \min_{t_k\in Y}\big( u(P_k \cap E_5) + w(P_k \setminus E_5)\big)=u(P_{v_{17}} \cap E_5) + w(P_{v_{17}} \setminus E_6)=41,$ $l(P_0 \cap E_5) + w(P_0 \setminus E_5)=32 \leq \min_{t_k\in Y}\big(l(P_k \cap P_0 \cap E_5)+ u(P_k \setminus P_0 \cap E_5) + w(P_k \setminus E_5)\big)=l(P_0\cap E_5) + w(P_0 \setminus E_5)=32$.Hence, $C_5=7$ is feasible. Then update $b:=5,k:=4.$ Then $b=a+1,k^*:=b=5$. 

\textbf{In Line 17: determine an optimal solution and its corresponding objective value.} 
From (\ref{fs-RIOVSPTBH}), \\
$e_j=e_{10}=(v_9,v_{10})$. 
We can obtain an optimal solution $$\tilde w:=(\textbf{9}, \textbf{18},     \textbf{12},     6,    \textbf{7},    12,      14,    19,   \textbf{ 17},    \textbf{3},    17,    \textbf{12},    38,    10,   14,     17)$$ and its corresponding objective value is  $C(\tilde w):=C_{k^*}=C_5=7.$ Then $\bar w$  is a pricing scheme to prevent arbitrage opportunities and also to ensure reliable emergency communication.

We next show the detailed process to solve the  problem \textbf{(2)} by \textbf{Algorithm \ref{alg-MCSPITBH}}.

\textbf{In Lines 1-2:} Construct the sequence $\{\alpha_{n^*}\}$  as above. 

\textbf{In Line 3:} Calculate $D(w):=\min_{t_k\in Y}w(P_k)=w(P_{v_6})=34$, Obtain $\hat w^{11}$ with $C_{11}=15$ from formula (\ref{os-MSPITBH}) and its relevant length of shortest root-leaf distance $D_{11}:=D(\hat w^{11})=u(P_{v_{17}})=50$. Obtain $\hat w^1$ with $C_1=1$ from formula (\ref{os-MSPITBH}) and its relevant length of shortest root-leaf distance $D_1:=D(\hat w^1)=\hat w^1(P_{v_{6}})=34$.

\textbf{In Lines 4-7:} judging special cases. $D_{11}=50>D:=39>D(w)=34$. Then the problem is feasible. 

\textbf{In Lines 9-22: a binary search method.}  \\
Initialization: $a:=1,b:=n^*=11,k^*:=0,k:=6$.

\textbf{In the first iteration},    obtain $\hat w^6$ with $C_6=8$ from formula (\ref{os-MSPITBH}) and its relevant length of shortest root-leaf distance $D_6 :=D(\hat w^6)=\hat w^6(P_{v_{17}})=41$.
Obtain $\hat w^{7}$ with $ C_{7}=9$ and its relevant length of shortest root-leaf distance $D_{7} := D(\hat w^{7})=\hat w^7(P_{v_{17}})=41$. Then $D_6>D:=39$. Update $b:=6, k:=3$.

\textbf{In the second iteration}, obtain $\hat w^3$ with $C_3=3$  from formula (\ref{os-MSPITBH}) and its relevant length of shortest root-leaf distance $D_3 :=D(\hat w^3)=\hat w^3(P_{v_{17}})=41$.
Obtain $\hat w^{4}$ with $ C_{4}=4$ and its relevant length of shortest root-leaf distance $D_{4} := D(\hat w^{4})=\hat w^4(P_{v_{17}})=41$. Then $D_3>D:=39$. Update $b:=3, k:=2$.

\textbf{In the third iteration}, obtain $\hat w^2$ with $C_2=2$ from formula (\ref{os-MSPITBH}) and its relevant length of shortest root-leaf distance $D_2 :=D(\hat w^2)=\hat w^2(P_{v_{6}})=40$.
Obtain $\hat w^{3}$ with $ C_{3}=3$ and its relevant length of shortest root-leaf distance $D_{3} := D(\hat w^{3})=\hat w^3(P_{v_{17}})=41$. Then $D_2>D:=39$. Update $b:=2, k:=1$.

\textbf{In the fourth iteration}, obtain $\hat w^1$ with $C_1=1$ from formula (\ref{os-MSPITBH}) and its relevant length of shortest root-leaf distance $D_1 :=D(\hat w^1)=\hat w^1P_{v_{6}})=40$.
Obtain $\hat w^{2}$ with $ C_{2}=2$ and its relevant length of shortest root-leaf distance $D_{2} := D(\hat w^{2})=\hat w^2(P_{v_{6}})=40$. Then $D_1>D:=39$. Update $b:=1, k:=1$.  Then $k^*:=a=1$ as $a=b=1$. 

\textbf{In Line 23: determine an optimal solution and its corresponding objective value.} 
Obtain $$\hat w^1=(7, 12,     8,     6,    \textbf{7},    12,      14,    19,    11,    11,    17,    9,    38,    10,   14,     17)$$ with $C_1=1$ from formula (\ref{os-MSPITBH}) and its relevant length of shortest root-leaf distance $$D_1 :=D(\hat w^1)=\hat w^1(P_{v_{6}})=40\geq D:=39.$$ Then $\bar w:=\hat w^1 $ is a pricing scheme just to prevent arbitrage opportunities.

\subsection{Numerical experiments}
We present the numerical experimental results for Algorithms \ref{alg-RIOVSPTBH} and \ref{alg-MCSPITBH} in Table \ref{table:ex}. We tested these algorithms on six randomly generated trees with vertex numbers ranging from 1000 to 5000.
We randomly generated the vectors $u$, $c$ and $w$ such that $0\leq w \leq u$ and $c > 0$. We generated $D$ for each tree based on $n$ with randomness. In this table the average, maximum and minimum CPU time are denoted by $T_i$, $T_i^{max}$ and $T_i^{min}$, $i=1,2$, respectively.

The experimental results demonstrate that both Algorithms \ref{alg-RIOVSPTBH} and \ref{alg-MCSPITBH} perform efficiently on large-scale trees, with Algorithm \( T_1 \) outperforming Algorithm \( T_2 \) in terms of CPU time since the sorting algorithm used in Algorithm \ref{alg-RIOVSPTBH} is more stable than the binary search in Algorithm \ref{alg-MCSPITBH}. 

\begin{table}[htbp!]
    \centering
    \caption{Performance of Algorithms \ref{alg-RIOVSPTBH} and \ref{alg-MCSPITBH}.}
    \label{table:ex} 
    \begin{tabular}{ccccc}\toprule   
        \textbf{Complexity} & \textbf{\( n \)} & \textbf{1000} & \textbf{3000} & \textbf{5000} \\ 
        \midrule
        \( O(n \log n) \) & \( T_1 \) & 0.3052 & 1.0427 & 1.8561 \\
                         & \( T_1^{\text{max}} \) & 0.5428 & 1.2613 & 2.1547 \\
                         & \( T_1^{\text{min}} \) & 0.1871 &  0.9701  & 1.4467 \\
        \midrule    
        \( O(n \log n) \) & \( T_2 \) & 0.5147 & 1.4928 & 2.5431 \\
                         & \( T_2^{\text{max}} \) & 0.6742 & 1.7624 & 2.9127 \\
                         & \( T_2^{\text{min}} \) & 0.3392 & 1.2691 & 2.1103 \\        
        \bottomrule
    \end{tabular}
\end{table}

\section{Conclusion and further research}
This paper addresses the restricted inverse optimal value problem for shortest paths under weighted bottleneck Hamming distance on trees \textbf{(RIOVSPT\textsubscript{BH})}. This optimization problem has significant applications in transportation and communication networks. Through rigorous analysis of its mathematical properties, we develop an efficient $O(n \log n)$ algorithm based on binary search methodology to solve \textbf{(RIOVSPT\textsubscript{BH})}. Additionally, we present a solution to the related \textbf{(MCSPIT\textsubscript{BH})} problem, achieving the same time complexity of $O(n \log n)$.

Our findings establish a foundation for future research directions. Potential extensions include investigating the restricted inverse optimal value problem on shortest paths for more complex graph structures, particularly series-parallel graphs and block graphs. Furthermore, the methodology developed here could be adapted to examine restricted inverse optimal value problems for other network performance metrics, such as maximum flow and minimum cost flow problems.

\end{document}